# Water flow in Carbon and Silicon Carbide nanotubes

Mara Cantoni, Edovardo Imalini

**ABSTRACT**

In this work the conduction of ion-water solution through two discrete bundles of armchair carbon and silicon carbide nanotubes, as useful membranes for water desalination, is studied. In order that studies on different types of nanotubes be comparable, the chiral vectors of C and Si-C nanotubes are selected as (7,7) and (5,5), respectively, so that a similar volume of fluid is investigated flowing through two similar dimension membranes. Different hydrostatic pressures are applied and the flow rates of water and ions are calculated through molecular dynamics simulations. Consequently, according to conductance of water per each nanotube, per nanosecond, it is perceived that at lower pressures (below 150 MPa) the Si-C nanotubes seem to be more applicable, while higher hydrostatic pressures make carbon nanotube membranes more suitable for water desalination.

## INTRODUCTION

Because of several different reasons, such as dealing with a worldwide shortage of rainfall and the necessity of replacing traditional water chlorination and using a reliable method to rub off only dangerous contaminants, there is a general tendency or special attention to find a useful way of seawater desalination and also filtration of harmful parasites from tap water. Some of the traditional methods of water purification are based on different membranes' applications, such as ionic membranes in electrodialysis which are semi permeable toward counter ions [1], that is, the repelling of equal ions and transferring opposite ones. The necessity of electrical energy along with producing waste water is a disadvantage of this method. The other applications belong to tiny-porous membranes through ultra-filtration and reverse osmosis processes.

The difference between these methods is that in former, the molecular dimension determines its transmission, whereas in latter, the water and organic molecules with a small molecular mass can pass through R.O membranes [2]. The role of nanotechnology in water processes contains three fields of quality development and filtration, sensing and pollution prevention. That is, using nano-based sensors to detect chemical pollutants accompanying with banning the others' way to enter water sources and also apply nanofiltration to desalinate and purify them. There is a wide range of nanoporous materials in environment, from biological systems to inorganic components and natural minerals, many of which have been applied as industrial membranes. According to development of observatory and manipulation tools, it is possible to design some porous membranes with precise dimensions and controllable physical and chemical characteristics. However, the complexity of biological systems prevents to understand the material transmission mechanism, selectivity and the role of hydrophobic coatings of pores. Therefore, it is preferred to use simple nanochannels to simulate ion permeation through membranes. Furthermore, these investigations will be used to design a new generation of molecular transmitters, being applied as water purification or bio-molecules dissociation. One of the good candidates of these nanochannels is carbon nanotube. According to some experimental [3] and theoretical [4,5,6] studies on carbon nanotubes, not only water permeation through narrow nanotubes is possible, but also the flow rates have been found to be unexpectedly high and independent of the pores' length. The observed properties of carbon nanotubes in these works, such as formation of water wires and the proton conduction along them, is similar to some biological pores like aquaporins [7,8]. While NaCl is dissolving in water, every positive and negative ion will be surrounded by water molecules through oxygen and hydrogen sides, respectively. Some studies show that the transmission of these solvated ions through membranes depends on nanotube's diameter and in some cases is so energetic and a membrane of (7,7) carbon nanotube is considered to be the best option to improve the flow rate of water [9].

In this work we aim to compare the conductance of water and NaCl ions, through a bundle of (7,7) carbon nanotubes under different hydrostatic pressures. We also try to replace carbon nanotubes by Si-C bundles and study the influence of other types of nanotube membranes on water and ion dissociation. In next sections, firstly we describe the manner of building initial models to do molecular dynamics simulations and then the results of our investigations are demonstrated and compared to make a general view of using two different types of nanotube bundles as desalination membranes.

## MOLECULAR DYNAMICS SIMULATIONS

The initial bundles are made of four carbon or silicon carbide nanotubes, packed hexagonally between two boxes containing both water molecules and sodium chloride ions. Because of different bond lengths of C-C (1.42 A$^o$) and Si-C (1.78 A$^o$), we build a bundle of (5,5) Si-C nanotubes to have a similar dimension to the former and also a comparable simulation results. The diameter and length of nanotubes are about 9.45 and 11.45 A$^o$, respectively. Each water box has dimensions of 27cm×21cm×30cm and contains two pairs of $Na^+$ and $Cl^-$.

The molecular dynamics simulations are done by NAMD software [10], using TIP3 model for water molecules. In order to study the interaction between nanotubes and fluid components, we calculate the total potential energy consisting of bonding (stretching, angle and dihedral) and long-term nonbonding interactions. Table 1 shows different force field parameters of carbon and silicon atoms.

The adjusted interval between integrations is assigned as 2 fs. The Nanotubes are fixed during the simulation and the Langevin dynamics is used to appoint a constant temperature in an NVT ensemble for 20 nanoseconds.

It is our aim to measure the transmission rate of water and salt ions through nanotubes, at different pressure differences. These hydrostatic pressure differences are employed according to Zhu et al. method [11,12-16], in which the oxygen atoms of waters are being pushed by a z-directed force. After a while, the velocity and consequently temperature of fluid increase rapidly and make water molecules vaporize. By coupling a Langevin thermostat to these atoms, we come over this problem. Thus, each atom is subjected to two forces, one as mentioned above and the other causes an opposite damping force. The pressure difference is calculated via $\Delta p = nf / A$, where n is the number of water molecules and A is the cross section area of the membrane. An approximately wide range of different forces are applied to water molecules and corresponding pressure differences are calculated for them (Table 2).

**Table 1.** Force field parameters for C and Si-C nanotubes

| Potential Parameters | CNT | | SiCNT |
|---|---|---|---|
| $V(bond) = K_b(b - b_0)^2$ | C-C | | C-Si |
| $b_0$ (A°) | 1.400 | | 1.783 |
| $K_b$ (kcal/mol A°²) | 469.000 | | 299.043 |
| $V(angle) = K_\theta(\theta - \theta_0)^2$ | C-C-C | | C-Si-C |
| $\theta_0$ (degree) | 35.00 | | 47.54 |
| $K_\theta$ (kcal/mol rad²) | 120.00 | | 119.00 |
| $V(dihedral) = K_\Phi[1 + \cos(n\Phi - \Phi_0)]$ | C-C-C-C | | Si-C-Si-C |
| $\Phi_0$ (degree) | 180.00 | | 180.00 |
| n* | 2 | | 3 |
| $K_\Phi$ (kcal/mol) | 3.100 | | 0.350 |
| $V(LJ) = \varepsilon_{i,j}[(R_{min,I,j}/r_{i,j})^{12} - 2(R_{min,I,j}/r_{i,j})^6]$ | C | C | Si |
| $\varepsilon$ (kcal/mol) | 0.0860 | 0.0559 | 0.5850 |
| $R_{min}/2$ (A°) | 1.910 | 1.910 | 1.904 |

*n is dihedral multiplicity

**Table 2.** The conductance of water molecules and ions at different applied pressures

| | CNT | | | | | SiCNT | | | | |
|---|---|---|---|---|---|---|---|---|---|---|
| Force kcal/molA° | $\Delta p$ Mpa | # H$_2$O | $Q_{fluid}$ cm³/s | # Na | # Cl | $\Delta p$ Mpa | # H$_2$O | $Q_{fluid}$ cm³/s | # Na | # Cl |
| 0.003 | 10.5 | 134 | 2.01×10⁻⁷ | 1 | 0 | 13.9 | 8 | 1.2×10⁻⁸ | 0 | 0 |
| 0.03 | 104 | 3064.5 | 4.59×10⁻⁶ | 7 | 1 | 139 | 8001.5 | 1.2×10⁻⁵ | 23.5 | 22 |
| 0.3 | 1048 | 145220 | 2.18×10⁻⁴ | 543 | 409.5 | 1396 | 171519.5 | 2.57×10⁻⁴ | 666 | 614.5 |

## RESULTS AND DISCUSSION

The analysis of simulation results are done by VMD package [13]. The number of water molecules inside each tube is calculated per nanosecond and the results for three chosen applied forces (related to different hydrostatic pressures) are shown in Figure 1. On the whole, the calculation of water flow through different nanotubes of carbon or silicon carbide shows more dependency of hydrostatic pressure for the former. At a pressure difference of about 10 Mpa, we can see that four carbon nanotubes almost alternatively become full or empty of water, but by an increase of hydrostatic pressure to 100 Mpa and more, all nanotubes are filled together (Figure 1, 2).

It is shown that at higher pressures, sodium ions can enter both nanotubes more easily than chlorides. This is because of a larger solvated layer around more massive Cl ions. Therefore, the dissociation of some water molecules, regarding to entrance of these ions is more energetic. Such a result has been mentioned in Corry's work too [9]. Our study represents that differences between positive and negative entering ions are smaller for silicon carbide bundles than carbon nanotubes (Table 2). However, by applying periodic boundary conditions, the movements of ions through bundles, don't accomplish any charge dissociation in the whole system.

The other noticeable result is the special arrangement of water molecules, as a double chain inside CNTs at three different applied pressures. However, the single file chain of water in an SiC bundle at pressures lower than 140 Mpa, turns into a double chain regarding to a ten times increase in applied pressure (Figure 3, up). As Corry has mentioned a slight diagonally elongation of (7,7) carbon nanotubes, we also observe that Si-C nanotubes are not excluded from this transformation while flowing of water molecules (Figure 3, down).

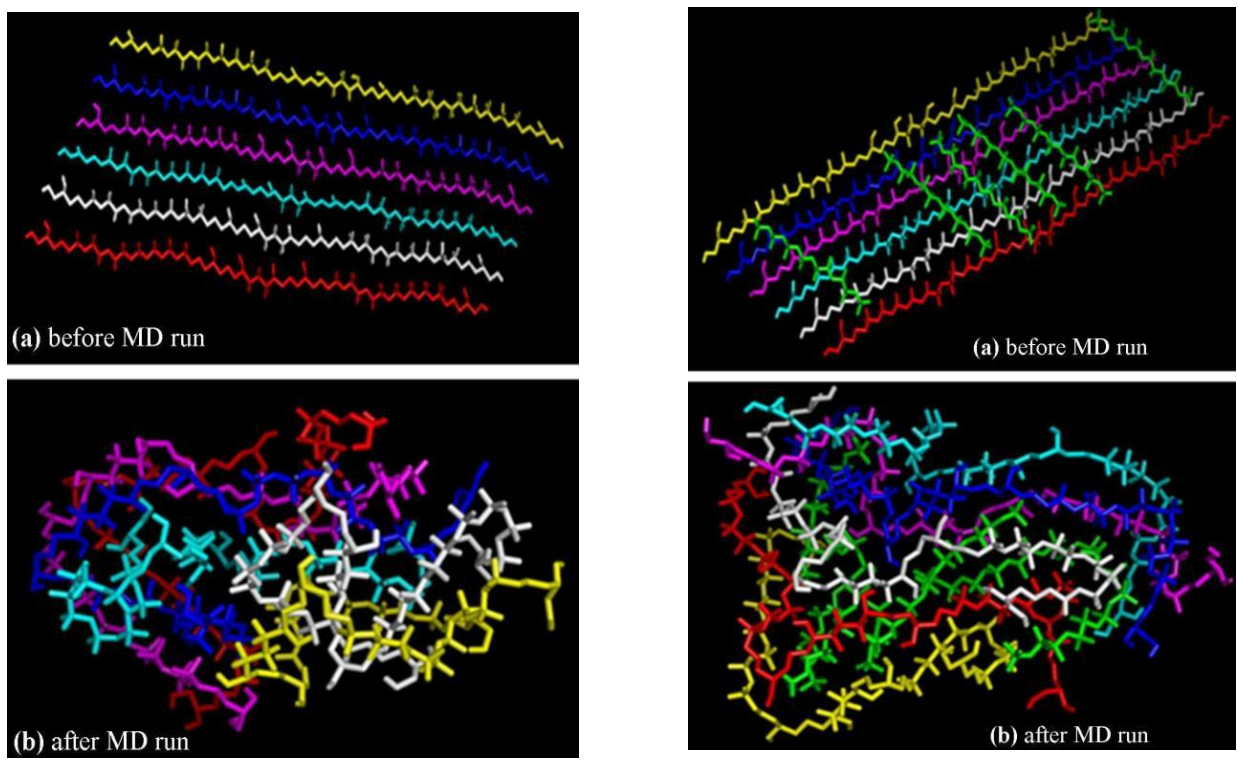

**Fig. 1.** The number of water filling four CNTs at different applied forces (kcal/molA), through 20 ns MD simulations

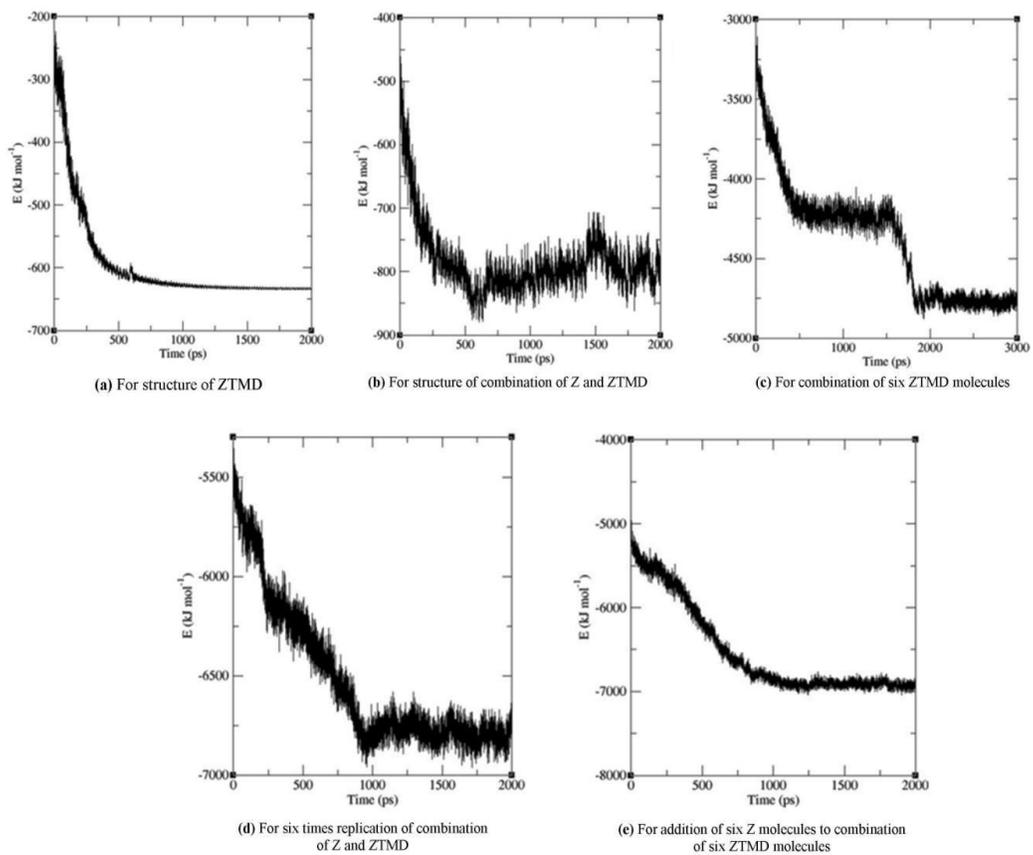

**Fig. 2.** The number of water filling four SiCNTs at different applied forces (kcal/molA), through 20 ns MD simulations

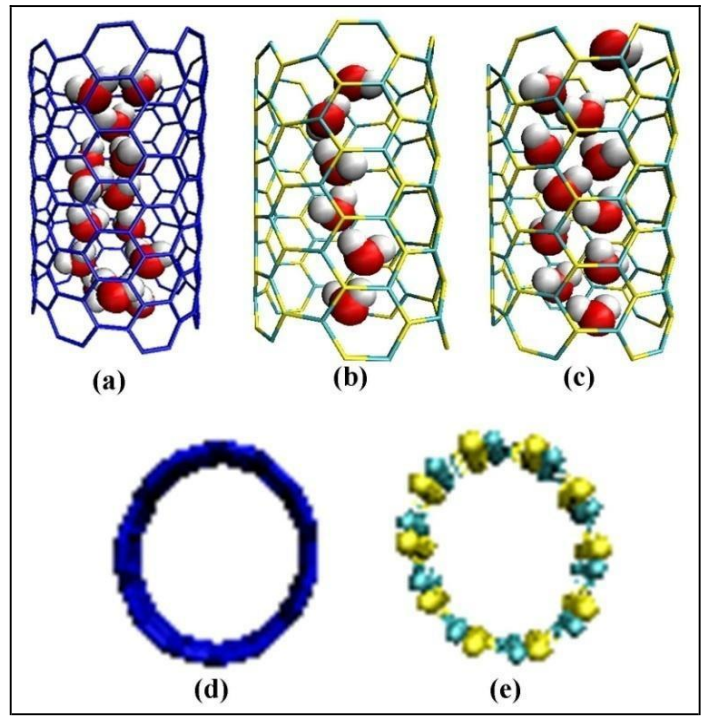

**Fig. 3.** Up: Water rearrangement inside studied CNTs (a) at different pressures and SiCNTs at pressures less than 140 MPa (b) and more (c). Down: The enlargement of carbon (d) and silicon carbide (e) nanotubes' wall.

Although, water molecules surround dissociated ions through solvated layer, each molecule may also be connected to four other ones through hydrogen bonds. According to diameter of each nanotube, some of these bonds should be broken while fluid transmission through bundles. Actually, such a connection between water molecules can be rearranged inside nanotubes and would lead to their sequential movements. The calculation of hydrogen bonds inside and at nearby distances to nanotubes, shows that at high pressure differences, the number of hydrogen bonds among water molecules trying to enter bundles increases noticeably. On the other hand, it is indicated that under same condition, the number of the molecules flowing through nanotubes increases by growth of applied pressures. Therefore, the reduction of hydrogen bonds' numbers doesn't make an important prevention against fluid conduction. We also demonstrated that most of hydrogen bonds should be broken while flowing through Si-C than C nanotubes (Figure 4). Figure 5 indicates the conductance of water molecules per nanosecond per carbon nanotube compared with silicon carbide nanobundles. It is observable that at lower pressures, the treatment of carbon nanotubes is better.

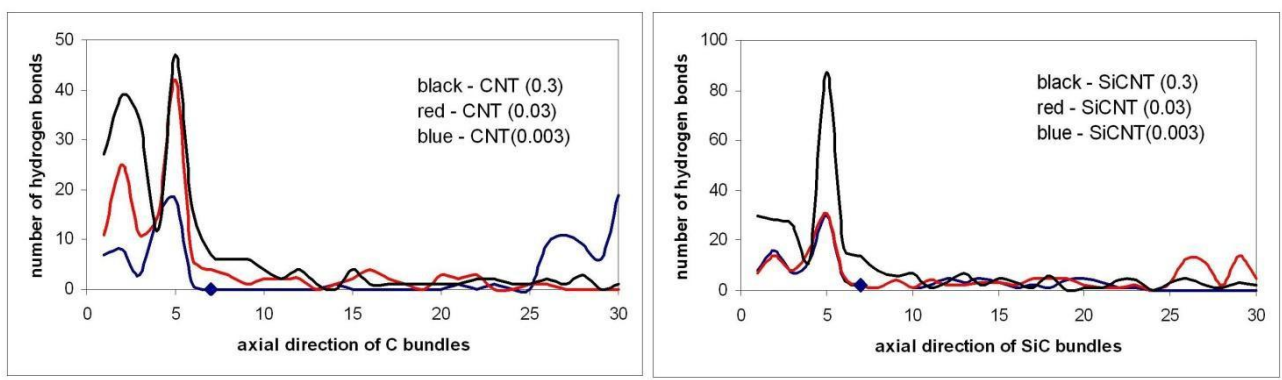

**Fig. 4.** The number of hydrogen bonds vs. carbon (left) and silicon carbide (right) bundles' direction at different applied forces (kcal/molA°). The entrance of membrane is indicated as point 7 of horizontal axis.

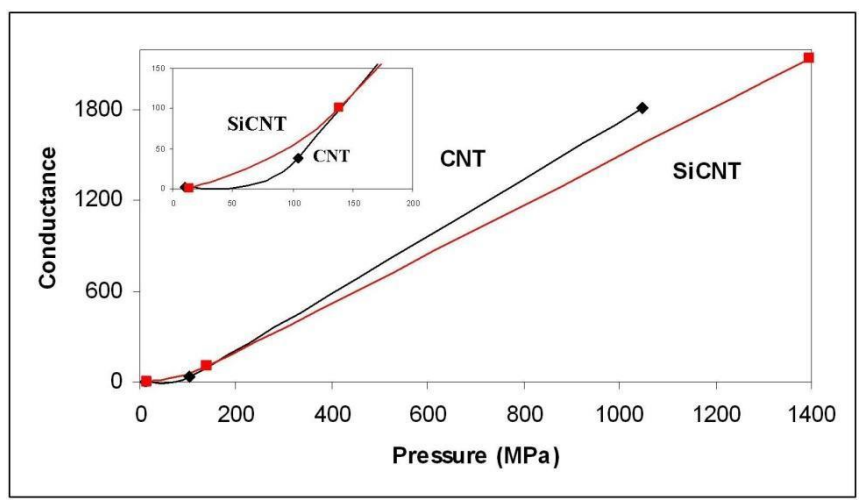

**Fig. 5.** The comparison of conductance of water molecules through CNTs and SiCNTs at different pressures

## CONCLUSION

Although carbon nanotubes are famed to have desirable properties, the understanding of similar applications of silicon carbide nanotubes is also noticeable. According to investigations of different characteristics of SiCNTs, they have even some superiority over carbon nanotubes via their more thermal stability, mechanical strength and also capacity of hydrogen adsorption [14]. Different types of carbon nanotubes have been studied as impermeable membranes for water molecules. However, we have shown that such investigations can be applied to study the conductance of an ionic fluid through silicon carbide nanotubes too.

The results for two different types of bundles, with similar dimensions showed some similarities, as dependence of water impermeability to applied pressure. However, according to the role of pressure in water conductance, it is concluded that at hydrostatic pressures below 150 Mpa, the conductance diagram for silicon carbide bundles has a sharper gradient indicating the better fluid conductance properties. However by increasing the pressure, this relation inverts in favour of carbon nanotube bundles (Figure 5). Regarding our simulations, it was also indicated that the dependence of CNT bundles to pressure changes seems to be more than SiCNTs. We have demonstrated the influence of water flowing through two studied bundles on the number of their hydrogen bonds and also on nanotubes' wall shape. Such results follow the later investigations done on different carbon nanotube membranes by various diameters.